\documentclass[twocolumn,showpacs,aps,prl]{revtex4}
\usepackage{graphicx}
\usepackage{epsf}
\newcommand{\etal}{{\it et al.}}

\begin{document}
	
\title{Shallow pockets and very strong coupling superconductivity in FeSe$_x$Te$_{1-x}$}
\author{Y. Lubashevsky, E. Lahoud, K. Chashka, D. Podolsky and  A. Kanigel}
\affiliation{Department of Physics, Technion, Haifa 32000, Israel}

\begin{abstract}
We measured the electronic-structure of FeSe$_x$Te$_{1-x}$ above and below T$_c$. In the normal state we find multiple bands with remarkably small values for the Fermi energy $\varepsilon_F$.  Yet,below T$_c$ we find a superconducting gap $\Delta$ that is comparable in size to $\varepsilon_F$, leading to a ratio $\Delta/\varepsilon_F\approx 0.5$ that is much larger than found in any previously studied superconductor.   We also observe an anomalous dispersion of the coherence peak which is very similar to the dispersion found in cold Fermi-gas experiments and which is consistent with the predictions of the BCS-BEC crossover theory.
\end{abstract}

\date{\today}
\maketitle

\begin{figure} 
\begin{center}
\includegraphics[width=8cm]{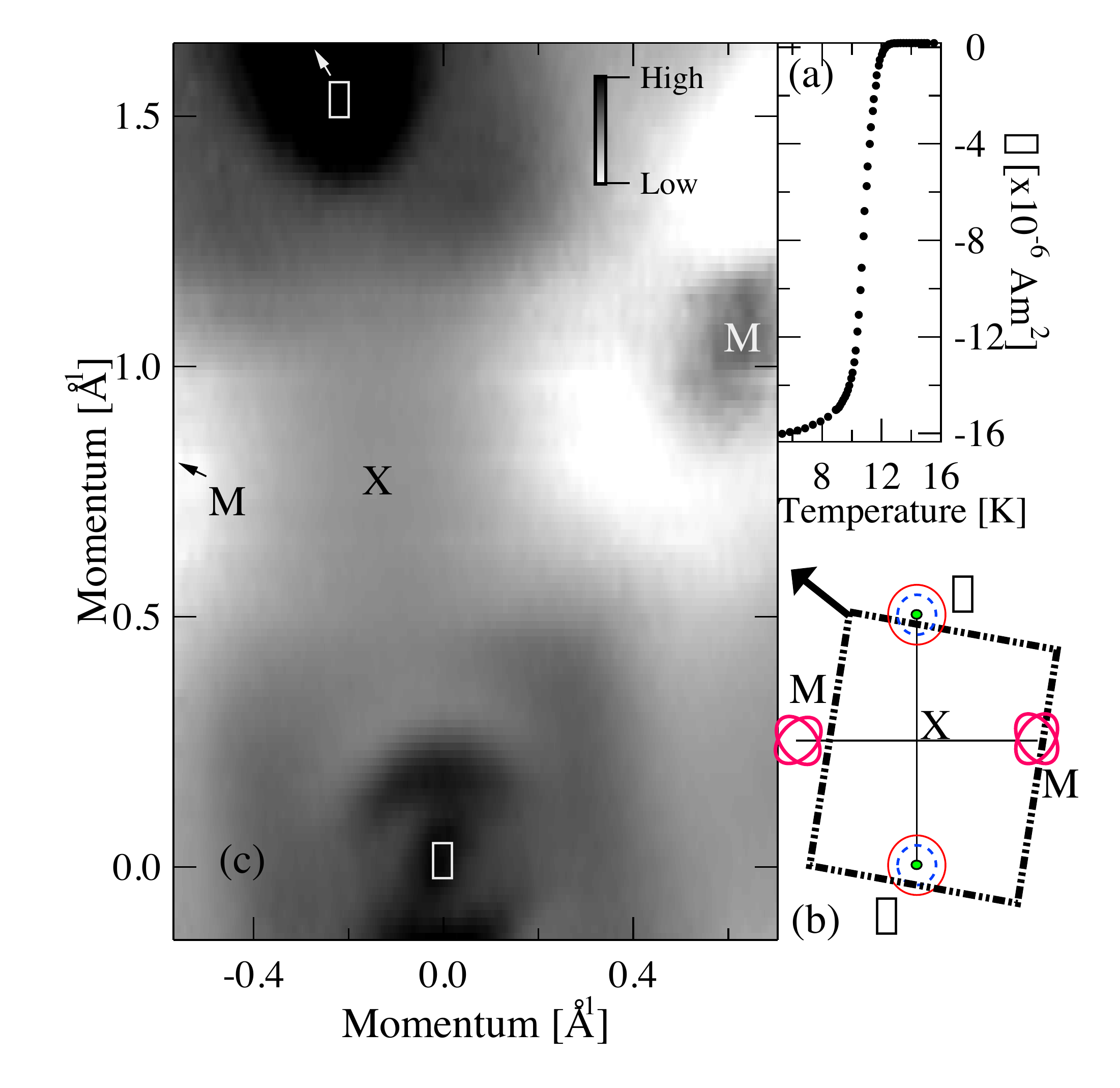}
\end{center}
\vspace{-1cm}
\caption{
 (a) Magnetic moment vs. temperature, showing a sharp SC transition at 12.5K (b) A sketch of the Brillouin zone of FeSe$_x$Te$_{1-x}$. (c) Intensity map at the Fermi level integrated over 4meV.}
\label{Fig1}
\end{figure}

The celebrated BCS theory, one of the greatest achievements in physics, is capable of explaining the behavior of metallic superconductors, but many believe that it is not adequate to deal with the newer high-temperature superconductors, the cuprates and the iron-based superconductors.

A possible extension of the BCS theory is provided by the BCS-BEC theory, which is based on the observation that the BCS wave-function can describe both a system of weakly interacting pairs and a BEC system of molecules of strongly-bounded fermions, as long as the chemical potential of the fermions is allowed to change sign as the interaction strength is increased. 
The behavior of the interacting fermionic system is governed by the interplay between the paring interaction represented by $\Delta$, the energy-gap, and the kinetic energy represented by the Fermi energy $\varepsilon_F$.

Cold fermi gasses present the best way to see directly the BCS-BEC crossover -- the Feshbach resonance offers a way to continuously change the interaction between the atoms\cite{Jin_crossover,Ketterle}. As the interaction strength is increased it is possible to observe the change in the chemical potential, $\mu$. It was shown using a novel RF-spectroscopy technique \cite{Jin_ARPES} that the momentum-distribution of the fermions changes in the way predicted by the BCS-BEC theory\cite{Eagles,Leggett,Mohit_review}. 

Up to now, a similar change in the momentum distribution has never been found in a solid material. 
The underdoped cuprates were considered the best candidates, since they have large gaps, of the order of 50 meV, and low carrier concentration, resulting in $\frac{\Delta}{\varepsilon_F} \sim 0.15$. It was suggested that many of the unconventional properties of the underdoped cuprates could be explained by the BCS-BEC crossover physics\cite{Levin_HighTc,Perali_BCS_BEC,Ranninger}, but spectroscopic evidence  of a change in $\mu$ was never found.

In this paper we show that the iron-chalcogenides are closer to the BCS-BEC crossover than the cuprates in terms of $\Delta/\varepsilon_F$ and that indeed below $T_c$ the electronic dispersion is consistent with the prediction of the BCS-BEC theory.  In a multi-band material we must introduce a Fermi energy $\varepsilon_F$ for each band, defined for electron-like (hole-like) bands as the energy of the highest occupied state relative to the bottom (top) of the band.  In particular, we find that the ratio $\Delta/\varepsilon_F$ is sizeable for {\em all} bands.  Furthermore, we find significant renormalization of the electronic dispersion below $T_c$, such that $\varepsilon_F$ for the hole-like bands becomes small or perhaps even negative, in close analogy to effects seen at the BCS-BEC crossover in cold Fermi gasses.

Superconductivity in the FeSe system (referred to as the ``11'' system) with a $T_c$ of 8K was discovered in 2008 \cite{Wu_PNAS}, later on it was shown that by replacing part of the Se with Te $T_c$ can be increased up to 15K \cite{Wu_EPL}. The iron-chalcogenides seem to be more strongly correlated compared to other iron-based superconductors (SC) \cite{Johnston}.

\begin{figure*}
\begin{center}
\includegraphics[width=19cm]{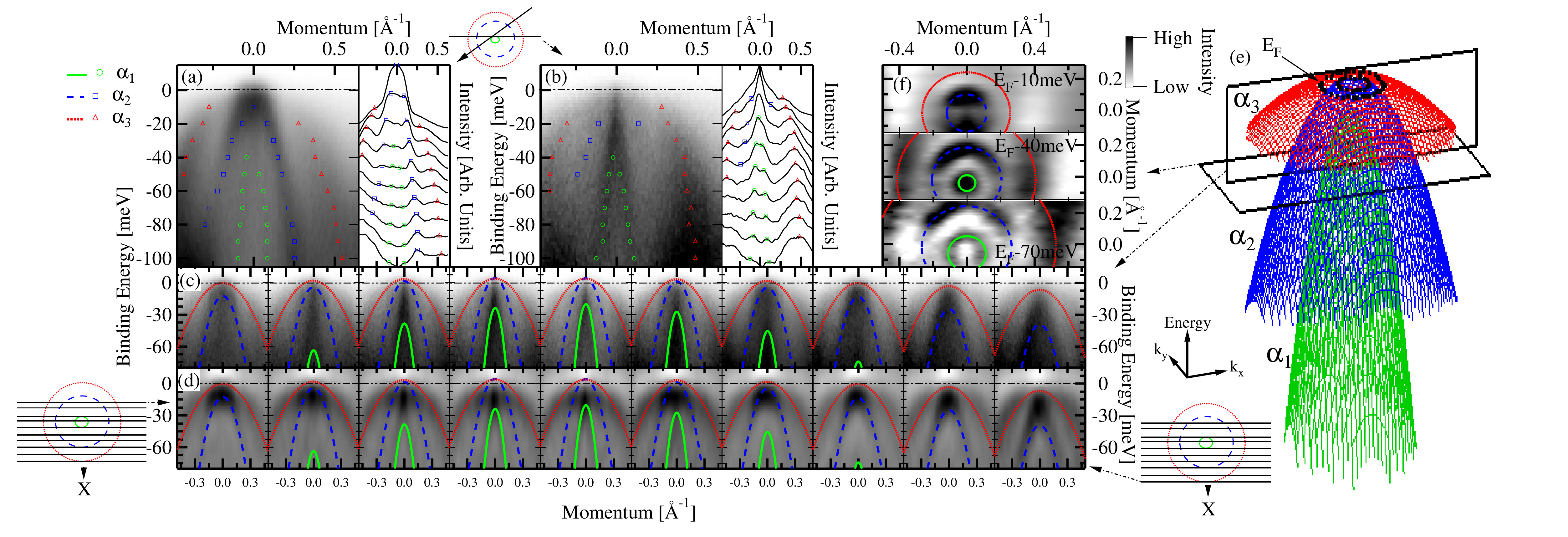}
\end{center}
\vspace{-1cm}
\caption{
$\Gamma$ point data:(a-b) ARPES data and the corresponding MDCs for a cut taken in the $\Gamma$-M and $\Gamma$-X  directions, respectively. (c) ARPES data for 10 different cuts taken parallel to the  $\Gamma$-X direction. (d) Second derivative in respect to energy of the data shown in (c). (e) The 2D fit-results, the 3 paraboloids are the best-fit to the data representing the 3 hole-like bands. We find that $\alpha_1$ is not crossing the Fermi-level. For both $\alpha_2$ and $\alpha_3$ we find small pockets with $\varepsilon_F$ =  4 $\pm$2.5 meV. (f) Intensity maps at 3 different binding energies.}
\label{Fig2}
\end{figure*}

All the iron-based superconductors are multi-band materials, which has a great impact on their transport properties and is believed to play a major role in the mechanism of superconductivity.
The band structure of the ``11'' system is similar to that of other iron-based superconductors. Calculations show hole-like bands at the center of the Brillouin zone ($\Gamma$ point), and electron-like bands around the M-point( see Fig.~\ref{Fig1}b). Combining the multi-band nature of the material and the strong-correlations, FeSe$_x$Te$_{1-x}$ is expected to have a rather complicated electronic-structure. 
 
We have performed Angle Resolved Photoemission Electron Spectroscopy (ARPES) measurements in order to map the band structure, both above and below $T_c$.
For the ARPES experiments we have grown high-quality single crystals of FeSe$_x$Te$_{1-x}$ using the ``self-flux'' method \cite{self_flux}. Energy Dispersive Spectroscopy shows a uniform composition with x=0.35. $T_c$ of these crystals is 12.5K, as found using SQUID magnetometry (see Fig \ref{Fig1}a). The transition is sharp ($\Delta$ T$_c<$ 1K) and the magnetic moment size indicates a large superconducting volume fraction.
The ARPES measurements were done at the Technion, using a Scienta R4000 and the HeI$_\alpha$ line (21.218eV) from an  He lamp. 
Crystals of typical size of $1\times 1$ mm$^2$ were cleaved in-situ at a pressure lower than $5\times 10^{-11}$ torr. The samples were cleaved at 8K, and measured below and above $T_c$.

We find a band structure similar to the one reported in previous ARPES work \cite{Tamai, Nakayama, Feng}. The symmetry and number of different bands agrees with the calculated band structure but the effective masses are strongly renormalized.

We start with the region around the $\Gamma$ point. The band structre above $T_c$, at 15K, can be seen in Fig \ref{Fig2}. 
The different orbital-character  of the bands leads to strong polarization and orientation dependence of the intensity \cite{Feng}. 
In panels (a) and (b) we show ARPES data and the corresponding Momentum Distribution Curves (MDC) taken with the detector parallel to $\Gamma$M and to the $\Gamma$X direction, respectively. Three bands can be identified, we use the notation; $\alpha_1$, $\alpha_2$ and $\alpha_3$ following Ref. \cite{Tamai}. $\alpha_1$ is more pronounced in panel (b) and $\alpha_2$ in panel (a). $\alpha_3$ is weak  but visible in both orientations. In order to extract the masses and the Fermi-crossing points of the different bands we did the following: we measured many cuts covering a rather large region around $\Gamma$. Some of these cuts can be seen in panel (c), the second derivative of the same cuts are shown in panel (d).  For each cut we used the MDCs to map out the dispersion. We used data from all the cuts to reconstruct the 2D band structure, fitting the data using three circular-paraboloids.  The results are shown in panel (e). 
The fit to the data is excellent, and it allows us to extract with a reasonably small uncertainty the Fermi-crossing points and the value of $\varepsilon_F$ for the various bands. The solid lines in panels (c) and (d) represents the fit results along different momentum cuts. 
In panel (f) we show intensity maps as a function of k$_x$ and k$_y$ at three different binding-energies: 10 meV, 40 meV and 70 meV. The solid lines are again the fit results now shown for the corresponding binding energies.  

We find that only $\alpha_2$ and $\alpha_3$ cross the Fermi-energy and create hole-pockets. $\alpha_1$ ends about 25meV below the Fermi-surface.  We find the following masses for the different bands: $m^{\star}_{\alpha_1}$ = 1.0$\pm$0.3 $m_e$, $m^{\star}_{\alpha_2}$ = 3.4$\pm$0.5 $m_e$ and $m^{\star}_{\alpha_3}$ = 14 $\pm 3$ $m_e$. These values are in good agreement with previous work \cite{Tamai}. The two bands that form the Fermi-surface are strongly renormalized in comparison to the calculated band-structure.  
We find two very shallow pockets; based on the fit we estimate $\varepsilon_F$ to be 4$\pm$2.5 meV for both pockets.  This is a surprising result: a metal with such a small Fermi-energy can't be considered a degenerate-Fermi gas even at room temperature. This is expected to lead to a non-trivial temperature dependence of various transport properties, as indeed was found for example for the Seebeck coefficient \cite{Thermo} and  Hall resistivity \cite{Hall}.
 
\begin{figure}
\begin{center}
\includegraphics[width=9.0cm]{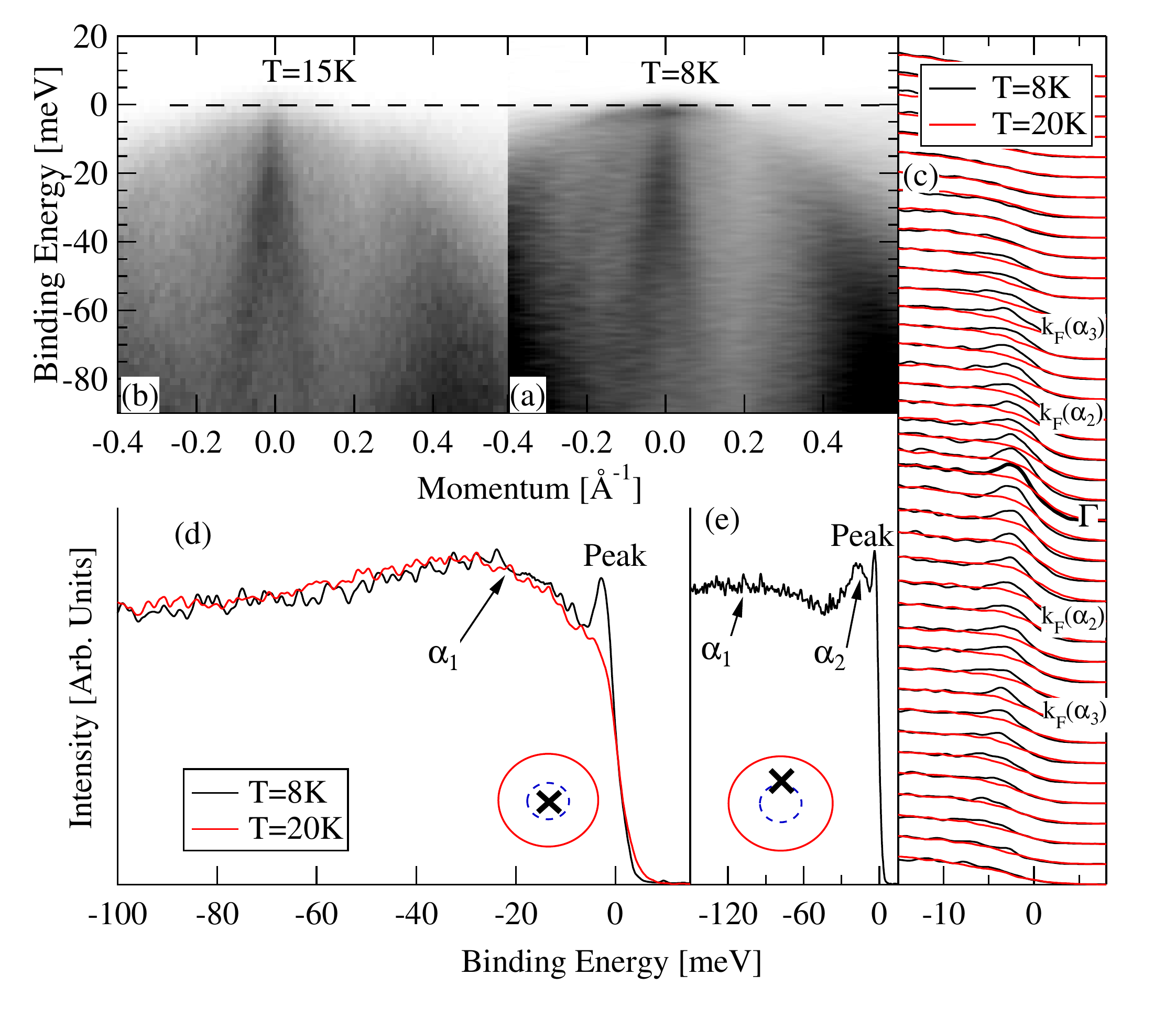}
\end{center}
\vspace{-1cm}
\caption{
Coherence peak: (a) Detector image for a cut going through the $\Gamma$ point taken at 8K. (b) ARPES data at the same momentum region taken at 15K. (c) A direct comparison of EDCs above and below $T_c$, the coherence peak is clearly visible. (d) A comparison of the EDC at the $\Gamma$ point above and below $T_c$. (e) EDC below $T_c$ at a momentum point between the two hole-pockets.}
\label{Fig3}
\end{figure}

When we lower the temperature below $T_c$, we find surprising results. 
In Fig \ref{Fig3}a we show the raw ARPES data of a cut going trough the $\Gamma$ point taken below $T_c$. In contrast to the data taken above $T_c$, Fig.~\ref{Fig3}b, one can clearly see the emergence of a coherence peak below $T_c$. In panel (c) there is a direct comparison of the Energy Distribution Curves (EDCs) measured at 20K and at 8K, the new peak is very sharp, its width is about 3meV and it is probably resolution limited. 
The coherence peak can be found over a large portion of the zone centered around the $\Gamma$ point, it is pronounced at that point although there in no Fermi-crossing point in the near vicinity. In panel (d) we compare the EDC at the $\Gamma$ point above and below $T_c$. There is an ``hump'' peaked around 25meV which its origin is the $\alpha_1$ band, the hump position and shape does not depend on temperature.  Below $T_c$ there is in addition a sharp peak.
For comparison we show in panel (e) the EDC below $T_c$ at a momentum point which lies between the two Fermi-pockets (see inset), here it is possible to identify both $\alpha_1$ and $\alpha_2$ in addition to the sharp coherence peak. It is possible to follow the coherence peak even beyond k$_F$ of the $\alpha_3$ band.

\begin{figure}
\begin{center}
\includegraphics[width=9.0cm]{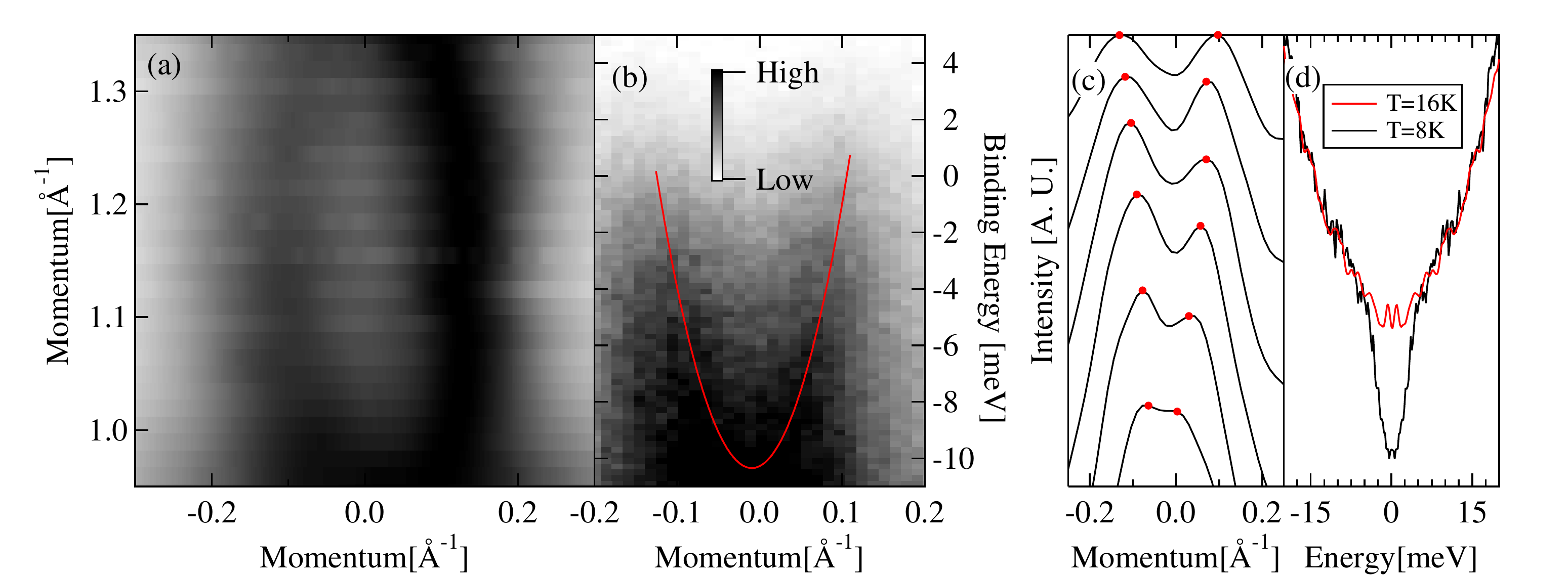}
\end{center}
\caption{M-point data: (a) An intensity map taken at the Fermi level . (b) A detector image for a cut going through the M-point. It is possible to see a faint electron-like dispersion (marked with a dashed line). (c) The MDCs showing the electron-like dispersion. (d) Symmetrized EDC at k$_F$ above and below $T_c$. A superconducting gap of about 5meV is opened.}
\label{Fig4}
\end{figure} 

 The data for the M-point region is not as clear. Figure \ref{Fig4}a shows the intensity map at the Fermi level, consisting of one oval pocket.  A second oval pocket, which is predicted by symmetry arguments to be perpendicular to the first, cannot be resolved.  A scan through the M point is shown in Fig.~\ref{Fig4}b, demonstrating a very shallow electron-like pocket (The MDCs are shown in pabel (c)). The bottom of this band is about 10 meV below the Fermi level. We find that band to have a mass of 4 $\pm$ 0.5 m$_e$, in reasonable agreement with band-structure calculations. Again, we find a very shallow pocket with a $\varepsilon_F$ similar to the one we find for the hole pockets. 
 In panel (d) we compare the EDC at $k_{F}$ measured above and below $T_c$. There is clearly a superconducting gap, but no coherence peaks. So far the size of the SC gap in the electron-pocket was not reported. The absence of  peaks and the large intensity tail coming from the hole-like band laying beneath the electron-like one, makes it difficult to measure the gap very precisely, but we estimate it to be of the order of 5meV.

\begin{figure}
\begin{center}
\includegraphics[width=8.0cm]{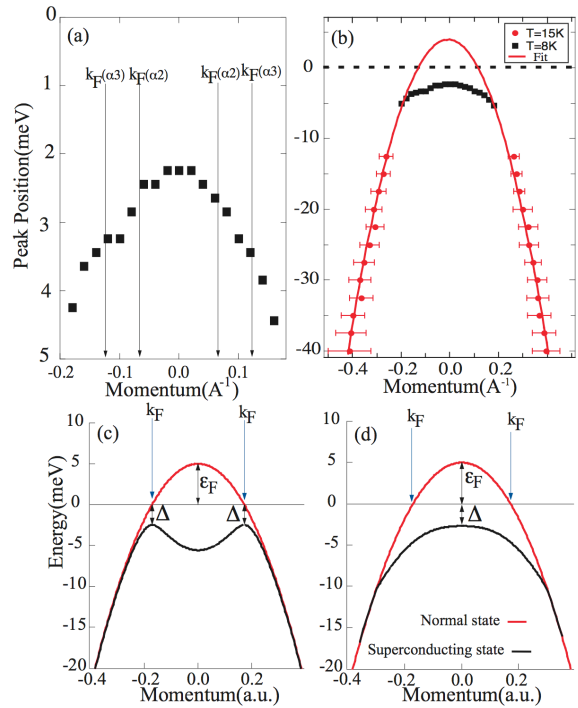}
\end{center}
\vspace{-1cm}
\caption{Dispersion below $T_c$ and the BCS-BEC crossover model: (a) Dispersion of the coherence-peak as extracted from the EDCs shown in Fig \ref{Fig3}c. (b) Dispersion of the $\alpha_3$ band. The red points represent MDC peak-positions, they are found to be the same above and below $T_c$. The black squares represent the coherence peak position. The red-line is from the 2D fit.
(c) A sketch of the quasiparticle-dispersion (Eq \ref{disp}) for the case $\tilde{\varepsilon}_F \sim \varepsilon_F$ (BCS limit). (d) A sketch of the quasiparticle-dispersion (Eq \ref{disp}) for the case $\tilde{\varepsilon}_F=0$ (BCS-BEC crossover regime). }
\label{Fig5}
\end{figure}

The coherence peak position disperses as one moves from the $\Gamma$ point, as shown in Fig \ref{Fig5}a. In Fig \ref{Fig5}b we show the dispersion of the $\alpha_3$ band and of the coherence peak. The dispersion of the $\alpha_3$ band was extracted  from the MDC peak positions.

 We can follow the band up to 12meV from $\varepsilon_F$, the same dispersion is found above and below $T_c$ within the error bars shown in the figure. The signal is very weak when approaching $\varepsilon_F$, and it is unlikely that this is attributed to a matrix-element effect since the intensity depends on the binding energy very strongly. This is a pseudogap, which is expected in the normal state when the interaction is strong enough \cite{Perali_PG}. The strong interactions broaden all the features making it impossible to follow the dispersion in the psedogap state up to $\varepsilon_F$.
 
 The dispersion below T$_c$ was extracted by following the peak-position in the EDCs. Remarkably, the minimal gap is found exactly at the $\Gamma$ point, and not around $k_F$.  In fact, nothing special happens at $k_F$, as can be seen in panel (a).  The coherence-peak dispersion reveals a very flat dispersion below $T_c$, in order to relate the dispersion of the coherence-peak to that of $\alpha_3$ we must conclude that the effective mass is strongly renormalized when the sample becomes superconducting.  

The gap size at the $\Gamma$ point is $\Delta=2.3 \pm 0.3$meV; the zero temperature gap should be somewhat larger.  Comparing that to $\varepsilon_F$, we get $\frac{\Delta}{\varepsilon_F} \sim 0.5$.  This is a large number, suggesting strong-coupling superconductivity in FeSe$_x$Te$_{1-x}$. For comparison $\frac{\Delta}{\varepsilon_F} \sim 0.1$ in optimally doped Bi2212. 
Before moving-on, let us summarize our findings: we find in the normal state that both the electron- and hole-like bands in Fe FeSe$_x$Te$_{1-x}$ form very shallow pockets, on the order of few meV, we anticipate that this might lead to an unusual electronic-structure below $T_c$ and indeed we find a coherence peak with a minimal gap at the $\Gamma$ point and not at $k_F$.

In the BCS theory, below $T_c$ the quasiparticle dispersion is given by:
\begin{equation}
\label{disp}
E_k=\pm \sqrt{\xi_k^2 + \Delta_k^2}
\end{equation} 
where $\Delta_k$ is the gap function and $\xi_k=\pm (\varepsilon_k-\tilde{\varepsilon}_F)$ is the band dispersion measured relative to a band-dependent shift $\tilde{\varepsilon}_F$.  Here, $\varepsilon_k=k^2/2m^{*}$, and the sign of $\xi_k$ depends on whether the band is electron-like or hole-like. For a single band, it is conventional to replace $\tilde{\varepsilon}_F$ by the chemical potential $\mu$ within the
crossover mean field theory.  In the current context, however, we prefer to use $\tilde{\varepsilon}_F$, since each band has its own Fermi energy, and in order to allow for the independent renormalization of the bands' Fermi energy, {\em e.g.} due to self-energy effects below $T_c$ \cite{CarlsonReddy}.

To find $\tilde{\varepsilon}_F$ and $\Delta_k$ one needs to solve the BCS gap and number equations self-consistently \cite{Schriefer}.  In the weak coupling limit $\tilde{\varepsilon}_F=\varepsilon_F$ to an excellent approximation and we get the Bogolyubov dispersion with the characteristic back-bending of the occupied branch. The dispersion is given by $E_k=-\sqrt{\varepsilon_k^2+\Delta^2}$, where the energy is measured relative to $\varepsilon_F$. The weak-coupling situation for a hole-like band is shown in Fig \ref{Fig5}c. In this case the minimum of the energy-gap occurs at $k_F$ and its value is $\Delta$. At k=0 the quasiparticle energy is $\sqrt{\varepsilon_F^2 + \Delta^2}$. This characteristic dispersion was measured using ARPES in the cuprates \cite{JC back bending} and in the pnictides \cite{Hong peak}. 

On the other hand, when the interaction is strong enough, there will be a significant shift in $\tilde{\varepsilon}_F$. As $\tilde{\varepsilon}_F$ becomes smaller, the position where the gap-minimum is found shifts away from the normal state k$_F$ towards k=0. In the extreme case where $\tilde{\varepsilon}_F$ reaches zero (or changes sign) the minimum of the gap moves to the $\Gamma$ point and its value becomes $E_\Gamma=-\sqrt{\tilde{\varepsilon}_F^2+\Delta^2}$ \cite{Mohit_review}. This quasiparticle dispersion as predicted by the model in the case where $\tilde{\varepsilon}_F$ is exactly 0 is shown in Fig.~\ref{Fig5}d. The similarity to our data, shown in Fig.~\ref{Fig5}b, is striking.

We find $\Delta/ \varepsilon_F=0.6 \pm 0.4$.  This estimate is based on the assumption that $\tilde{\varepsilon}_F$ is close to zero.  Although we cannot determine $\Delta$ and $\tilde{\varepsilon}_F$ independently, our data is not consistent with a very negative $\tilde{\varepsilon}_F$, as that would lead to a much larger binding-energy at the $\Gamma$ point than the normal-state $\varepsilon_F$. On the other hand, if the change in $\tilde{\varepsilon}_F$ were small, the minimum of the binding-energy below $T_c$ would be at $k_F$ instead of at the $\Gamma$ point.  This indicates that our sample does not lie deep in either the BCS or BEC regimes but lies instead in the crossover regime.

Note that in cold fermion gasses, the number of particles is kept constant and the BEC-BCS crossover is reached by tuning the interaction strength by going through a Feshbach resonance.  By comparison, in ARPES experiments, the chemical potential is tethered to the reference sample, gold in our case.  Although we are not free to change the chemical potential, we find that below $T_c$ there is substantial renormalization of the electronic dispersion, forcing one of the hole bands to move across the chemical potential. 

In its simplest form, the BCS-BEC crossover theory is in good qualitative agreement with our data, although it doesn't give a full quantitative description.   In particular, the mean-field theory predicts that for a 2D system the gap must be twice as large as $\varepsilon_F$ in order for $\tilde{\varepsilon}_F$ to change sign upon crossing $T_c$\cite{2D_BCS_BEC}.   To get quantitative agreement one may need to include multiple bands, momentum-dependent interactions, and self-energy corrections, which play an important role in renormalizing the band structure.

There may be alternative ways to interpret our ARPES data. The line-shape with the sharp peak at the gap edge below $T_c$ and the broader hump at lower energies (see Fig \ref{Fig3}d) is very similar to the ARPES line shape found in the underdoped cuprates. In the cuprates the line shape and the anomalous dispersion were explained as a result of the appearance of a resonance-mode seen in neutron data \cite{Norman}. In Bi2212 the peak is found even as one moves away from $k_F$, which is consistent with our finding.  Recently a resonance-mode was found in FeSe$_x$Te$_{1-x}$ with an energy of 6.5meV \cite{Qiu_res_mode}, this energy seems to be a bit too large since the resonance-mode model requires $\omega_{res} < 2 \Delta$. It is difficult to see a reason for the resonance-mode interaction with the electrons to shift the minimum of the excaitation-gap from k$_F$ to zero, but we can't rule out the possibility that the resonance-mode plays a role in shaping the electronic dispersion below $T_c$. 

To summarize, we measured in detail the band structure of FeSe$_x$Te$_{1-x}$, we find both the electron and the hole-pockets to be very shallow with $\varepsilon_F$ of the order of few meV.  Below $T_c$ we find a SC gap which its size is a substantial fraction of $\varepsilon_F$, suggesting that FeSe$_x$Te$_{1-x}$ is on the verge of the BCS-BEC crossover limit. The anomalous dispersion of the coherence peak is in agreement with such a suggestion. 

We acknowledge useful discussions with Mohit Randeria. 
This research was supported by the Israeli Science Foundation and by the E. and J. Bishop research fund.

\end{document}